\documentclass{ws-procs975x65}

\newcommand{\be}{\begin{equation}}
\newcommand{\ee}{\end{equation}}
\newcommand{\beq}{\begin{eqnarray}}
\newcommand{\eeq}{\end{eqnarray}}

\begin{document}

\title{THE DYNAMICS OF SCALAR-TENSOR COSMOLOGY FROM RS TWO-BRANE MODEL}

\author{P. KUUSK$^*$, L. J\"ARV, and M. SAAL}
\address{Institute of Physics, University of Tartu,\\
Riia 142, 51014, Tartu, Estonia\\
$^*$E-mail: piret.kuusk@ut.ee}


\begin{abstract}
We consider  Randall-Sundrum two-brane cosmological model
in the low energy gradient expansion approximation by Kanno and Soda.
It is a scalar-tensor theory with a specific coupling function. 
We find a first integral of equations for the A-brane metric and
estimate constraints for the dark radiation term.
We perform a complementary analysis of the dynamics of the scalar 
field (radion) using phase space methods and examine 
convergence towards the limit of general relativity.
We find that it is possible to stabilize the radion at a finite value 
with suitable negative matter densities on the B-brane.
\end{abstract}


\bodymatter

\section{Introduction}

Following Kanno and Soda \cite{ks1}, we consider 
the Randall-Sundrum type I cosmological scenario \cite{rs1} 
with two branes (A and B) moving in a 5-dimensional bulk.
Both branes are taken to be homogeneous and isotropic 
and supporting energy-momentum tensors of
a perfect barotropic fluid with barotropic index $\Gamma$
($p= (\Gamma - 1) \rho$) on the A-brane 
(identified with our visible Universe) 
and $\Gamma^{^B}$ on the B-brane;
for simplicity we assume $\Gamma = \Gamma^{^B}$.

The field equations of the effective 4-dimensional theory
obtained by Kanno and Soda \cite{ks1} are
the equations of a scalar-tensor theory
with one scalar field $\Psi$ (interpreted as a radion) 
with a specific coupling function 
$\omega (\Psi) = {3 \Psi}/{(2 (1 - \Psi))}$
which describes the proper distance between the branes.
The cosmology of this model was first studied by 
Kanno et al. \cite{kss} and later by us \cite{iii, jks}.

\section{Field equations and analytic solutions}

The field equations \cite{ks1} on the A-brane for the 
Friedmann-Lema\^itre-Robertson-Walker (FLRW) line element 
$ds^2 = -dt^2 + a^2(t) [(1-kr^2)^{-1} \ dr^2 + r^2 d\Omega^2]$ 
and  perfect fluid matter on both branes read 
($H \equiv \dot{a}/a$) 
\beq 
\label{00}
H^2 =  
- H \frac{\dot \Psi}{\Psi} 
+ \frac{1}{4} \frac{\dot \Psi^2}{\Psi (1 - \Psi)} + \frac{\kappa^2}{3} 
\frac{V}{\Psi} 
+ \frac{\kappa^2}{3} \frac{\rho}{\Psi} + \frac{\kappa^2}{3} \frac{(1- 
\Psi)^2}{\Psi} \rho^{_B} 
- \frac{k }{a^2} , \\ \nonumber \\ 
\label{mn}
2 \dot{H} + 3H^2 = 
-\frac{\kappa^2}{\Psi} p - \frac{\kappa^2}{\Psi} (1-\Psi)^2 p^{_B}  + 
\frac{\kappa^2}{\Psi} V 
- \frac{\ddot{\Psi}}{\Psi} - 2 H \frac{\dot{\Psi}}{\Psi} - \frac{3}{4} 
\frac{\dot{\Psi}^2}{\Psi(1-\Psi)} 
- \frac{k }{a^2} , 
\label{deq}
\eeq
\beq
\ddot \Psi &=& - 3 H \dot \Psi - \frac{1}{2} \frac{\dot \Psi^2}{(1-\Psi)} 
+ \frac{2}{3} \kappa^2 \ \left(2 V  - \Psi \frac{dV}{d\Psi}\right)
\left( 1 - \Psi \right)
  \nonumber \\ 
&& + \frac{\kappa^2}{3}(1-\Psi) \ ( \rho - 3p) 
+ \frac{ \kappa^2}{3}(1-\Psi)^2   ( \rho^{_B} - 3 p^{_B} ) , 
\eeq
they reduce to general relativity when $\Psi \rightarrow 1$, 
$\dot{\Psi} \rightarrow 0$.
The conservation laws as measured by an A-brane observer, 
$\dot \rho + 3 H \Gamma \rho = 0$ \,,
$\dot{\rho}^{_B} + 3 H \Gamma \rho^{_B} - (3 \dot{\Psi} \Gamma \rho^{_B})/ (2 (1-\Psi)) = 0$ ,
imply a relation between the energy densities on the A- and B-brane 
\beq \label{rhoAB}
\rho^{_B} = \rho^{_B}_{0} \left({\frac{1-\Psi}{1-\Psi_{0}}} \right)^{-\frac{3}{2} \Gamma} 
\left( \frac{\rho}{\rho_{0}} \right) \, .
\eeq
The dynamical equation for $H$ decouples from the scalar field and B-brane matter 
\cite{iii, jks} due to the specific form of the coupling function $\omega(\Psi)$  
and its first integral reads
\be \label{H^2}
H^2 = 
\frac{\kappa^2}{3} \, \sigma + 
\frac{\kappa^2}{3} \rho_{0} 
\left(\frac{a}{a_{0}}\right)^{-3 \Gamma} 
- \frac{k }{a^2} + 
\frac{\kappa^2 \,C}{3} \left(\frac{a}{a_0}\right)^{-4} \,,
\ee 
which is a Friedmann equation with a dark radiation term, $C$.
Comparison with recent results of light element abundances, BBN, and CMB
observations constrain the dark radiation term \cite{jks} 
($\rho_{0}$ is the radiation energy density),
\beq
-0.054 \leq  \frac{C}{\rho_0} \leq 0.138  \,. 
\eeq

Analytic solutions are found \cite{jks} for the flat Universe scale factor in the 
case of 
cosmological constant ($\Gamma= 0$),
radiation ($\Gamma = 4/3$),
dust ($\Gamma=1$),
cosmological constant and radiation,
and for the scalar field in the case of
radiation ($\Gamma = 4/3$),
cosmological constant and radiation (to be expressed in terms of elliptic functions).

\section{The dynamics of the scalar field}

Defining a new time variable \cite{dn} ,
$dp=h_c \, dt , \quad h_c = H + \frac{\dot{\Psi}}{2\Psi}$ ,
it is possible to derive a decoupled ``master equation'' for the scalar field 
(here $V = 0$, $k=0, \frac{df}{dp} \equiv f'$), 
\beq \label{master equation}
8 (1-\Psi)\frac{\Psi''}{\Psi} &-& 3(2-\Gamma)\left( \frac{\Psi'}{\Psi} \right)^3
-2 \Bigl( (4-6\Psi) - (4-3\Gamma)(1-\Psi) W(\Psi) \Bigr) 
\left(\frac{\Psi'}{\Psi}\right)^2  \nonumber \\
&+& 12 (2-\Gamma)(1-\Psi)\frac{\Psi'}{\Psi}
- 8(4-3\Gamma)(1-\Psi)^2 W(\Psi) = 0,
\eeq 
where 
\be
W(\Psi) = \frac{ 1+(1-\Psi) \  \frac{\ \rho^{_B}_0}{\rho_0} \left(
\frac{1-\Psi}{\ 1-\Psi_0} \right)^{-\frac{3}{2}\Gamma}}
{1+(1-\Psi)^2 \, \frac{\ \rho^{_B}_0}{\rho_0} \left( \frac{1-\Psi}{\ 1-\Psi_0}
\right)^{-\frac{3}{2}\Gamma}}  
\quad \rm{or} \quad 
W(\Psi) = \frac{1+(1-\Psi) \  \frac{\ \sigma^{_B}}{\sigma}}{1+(1-\Psi)^2 \,
\frac{\ \sigma^{_B}}{\sigma}}  ,
\ee
corresponding to the case when $\sigma = 0, \sigma^{_B} = 0,\rho \neq 0, \rho^{_B} \neq 0$ 
and  to the case when $\sigma \neq 0, \sigma^{_B} \neq 0, \rho = 0, \rho^{_B} =0$, respectively.
Phase portraits are found \cite{jks} depending on the values of constraints involved (Fig. ~\ref{plots}).

\def\figsubcap#1{\par\noindent\centering\footnotesize(#1)}
\begin{figure}[t]%
\begin{center}
 \parbox{2.1in}{\epsfig{figure=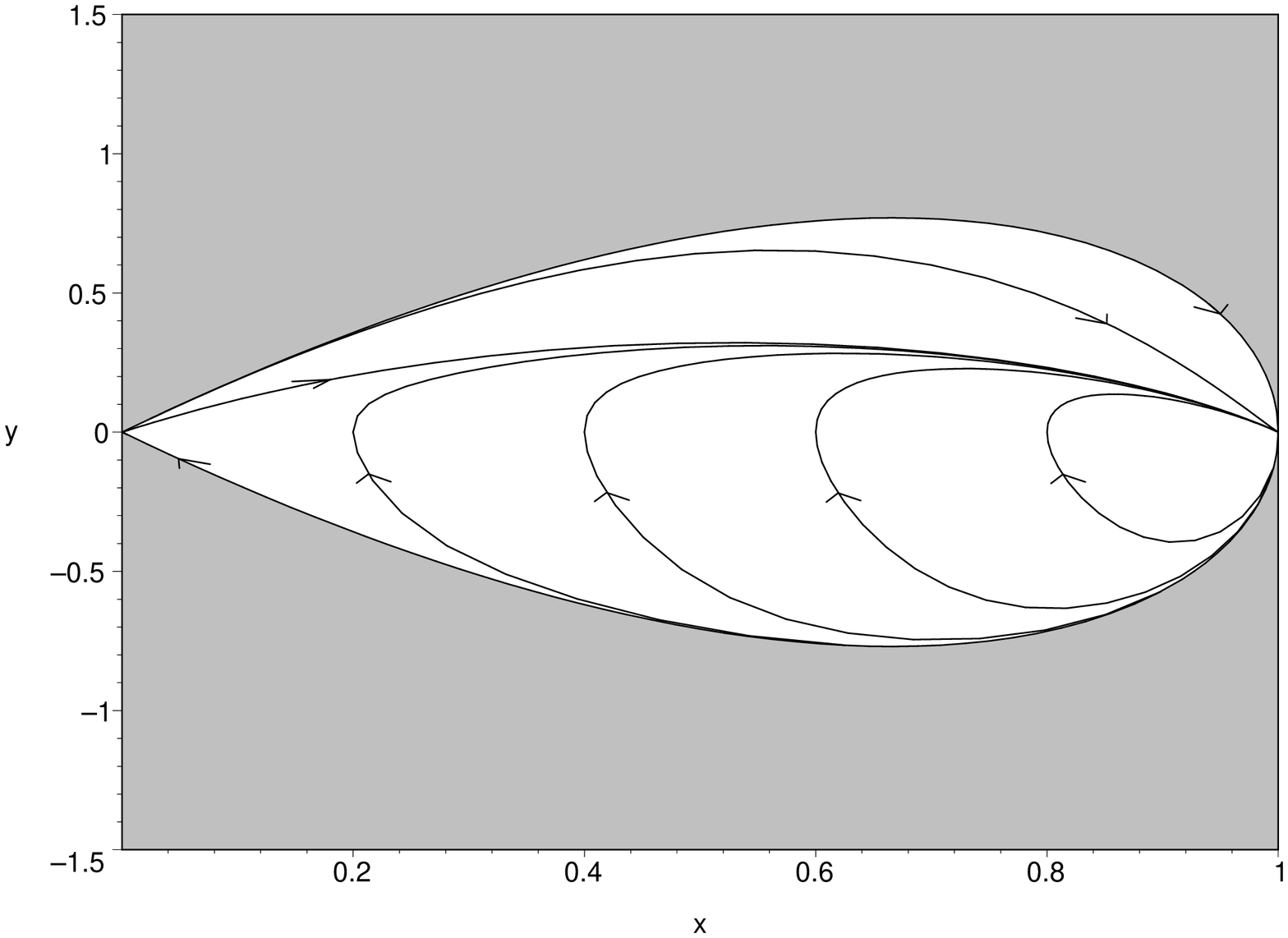,angle=-90, width=6cm}
 \figsubcap{a}}
 \hspace*{4pt}
 \parbox{2.1in}{\epsfig{figure=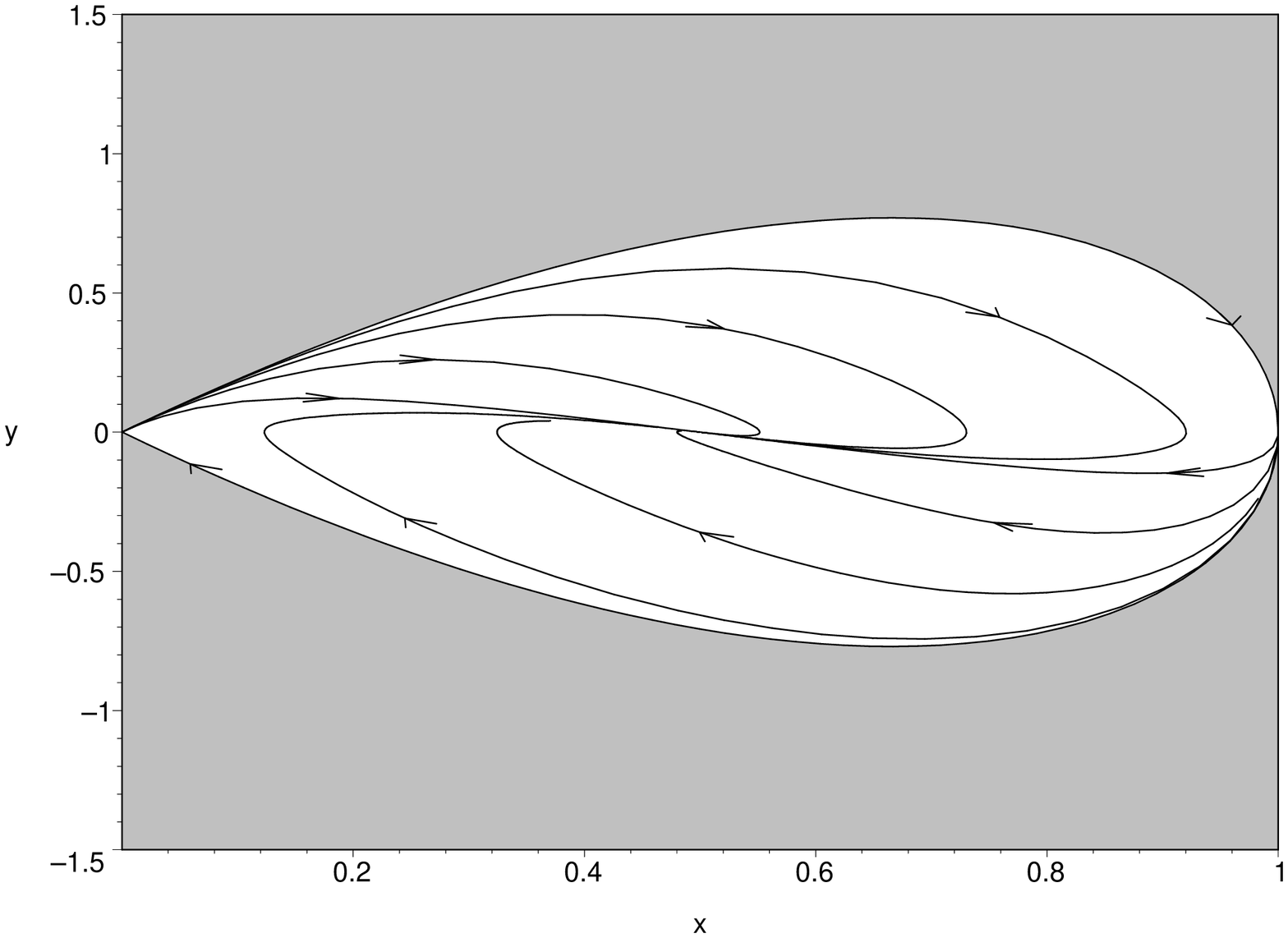,angle=-90, width=6cm}
 \figsubcap{b}}
\caption{Phase portraits ($x=\Psi(p)$, $y=\Psi'(p)$) (a) for cosmological 
constants: $\sigma = 1, \sigma^{_B} = -0.5$ and (b) for dust: 
$\rho_0 = 0.5, \rho^{_B}_0 = -1, \Psi_0 = 0.5$.}
\label{plots}
\end{center}
\end{figure}

All trajectories are constrained to be in the physically allowed region of the 
phase space determined by Friedmann equation (\ref{00}).
Figure 1a for a cosmological constant dominated Universe contains a saddle 
point ($\Psi = 0$, $\Psi' = 0$) and a spiralling attractor corresponding 
to general relativity ($\Psi = 1$, $\Psi' = 0$). 
Figure 1b for a dust dominated Universe with $\rho^{_B}_{0} < 0$ 
contains two saddle points, ($\Psi = 0$, $\Psi' = 0$) and 
($\Psi = 1$, $\Psi' = 0$), and also an attractor   
($\Psi = 1 - {\rho^{_B}_{0}}^{2} (1-\Psi)^3/ \rho_{0}^2$, $\Psi' = 0$)
which can stabilize the branes in a position that does not
correspond to general relativity on the A-brane.

\section{Summary}

We have considered a braneworld inspired scalar-tensor cosmology with
a specific coupling function, cosmological constant and 
perfect fluid matter on both branes.
The first integral of equations for the metric tensor of the A-brane 
contains the dark radiation term.
Phase portraits of the scalar field reveal fixed points and allow us
to find late time fates for different cosmological models.
The solutions may approach general relativity ($\Psi = 1$), 
or go to brane collision ($\Psi = 0$), depending on the initial conditions.
There are also additional fixed points in between the two extremes:
a saddle for cosmological constants and an attractor for dust (Fig. ~\ref{plots}(b)).


\begin{thebibliography}{9}

\bibitem{ks1}
S. Kanno and J. Soda, 
{\em Phys. Rev. D} {\bf 66}, 083506 (2002), 
[hep-th/0207029].

\bibitem{rs1}
L. Randall and  R. Sundrum,  
{\em Phys. Rev. Lett.} {\bf 83}, 3370 (1999),  
[hep-ph/9905221].

\bibitem{kss}
S. Kanno, M. Sasaki, and J. Soda, 
{\em Prog. Theor. Phys.} {\bf 109}, 357 (2003), 
[hep-th/0210250].

\bibitem{iii}
P. Kuusk and M. Saal, 
{\em Gen. Rel. Grav.}   {\bf 36}, 1001 (2004), 
[gr-qc/0309084].

\bibitem{jks}
L. J\"arv, P. Kuusk, and M. Saal,
{\em Phys. Rev. D} {\bf 75}, 023505 (2007), 
[gr-qc/0608109]. 

\bibitem{dn}
T. Damour and K. Nordtvedt, 
{\em Phys Rev. D} {\bf 48}, 3436 (1993).

\end{thebibliography}
\end{document}